\documentstyle[11pt]{article}

\begin{document}
%

\begin{center}
{\large \bf The Role of Solar Neutrinos in the Jupiter} \vskip.5cm

Valery Burov and W-Y. Pauchy Hwang\footnote{Correspondence Author;
 Email: wyhwang@phys.ntu.edu.tw} \\
{\em The Leung Center for Cosmology and Particle Astrophysics, \\
Institute of Astrophysics, Center for Theoretical Sciences,\\
and Department of Physics, National Taiwan University,
     Taipei 106, Taiwan}
\vskip.2cm


{\small(May 26, 2008)}
\end{center}

\begin{abstract}
Judging from the fact that the planet Jupiter is bigger in size
than the Earth by $10^3$ while is smaller than the Sun by $10^3$
and that the average distance of the Jupiter from the Sun is
$5.203 a.u.$, the solar neutrinos, when encounter the Jupiter, may
have some accumulating effects bigger than on the Earth. We begin
by estimating how much energy/power carried by solar neutrinos get
transferred by this unique process, to confirm that solar
neutrinos, despite of their feeble neutral weak current
interactions, might deposit enough energy in the Jupiter. We also
speculate on the other remarkable effects.

{\parindent=0pt PACS Indices: 96.40.Tv (Neutrinos and muons);
96.30.Kf (Jupiter); 95.85.Ry (Neutrino, etc.).}
\end{abstract}

\section{Introduction}

When we look at the eight major planets of our solar system, we
cannot stop being curious by many questions - and many puzzles to
ask. From inside out, the Mercury, the Venus, the Earth, and the
Mars, these might look like the earthlings, and then the Jupiter
and the Saturn might be mini-Suns, the other two we don't really
know. The fact that all eight planets fall in the same plane with
the same direction might indicate that they might form in similar
or related times. For those earthlings, the Mercury, the Venus,
the Earth, and the Mars, why only on Earth are there living
things?

The Sun provides the energy resources of all kind - the light, the
electromagnetic waves of all frequencies, the neutrinos, and the
cosmic-ray particles of all kinds - the main provider of the
extraterrestrial origin. Besides the light, solar neutrinos, which
come from the nuclear reactions in the core of the Sun, also carry
away a huge amount of energy. Differing from the light, solar
neutrinos, once produced, would travel up to the astronomical
distance without suffering second (weak) interactions.

Solar neutrinos and all other neutrinos would be the thing that
would "shine" the dark world, after all the lights cease to ignite
- another life of the Universe if the Universe ceases to expand or
start to contract. It would be a lot more interesting to look at
neutrinos and antineutrinos more intimately.

\section{Solar Neutrinos}

When the Sun is shining on us, a significant amount of the solar
energy get carried away by neutrinos. Solar neutrinos are elusive
because they only participate weak interactions - so almost all of
them pass away by us without being noticed. In fact, solar
neutrinos are even more elusive than, e.g., antineutrinos because
charged weak interactions do not operate between solar neutrinos
and the ordinary low$-Z$ matter, i.e. break-up of light nuclei by
solar neutrinos being negligible - they are made of from the
matter rather than the antimatter.

Solar neutrinos come from the most important reactions in the
so-called $pp$-I chain,\cite{Commins}

\begin{equation}
p+p \to D+e^+ +\nu_e, \qquad (E_\nu^{max}=0.42\,MeV: \,
\phi_\nu=6.0\times 10^{10}cm^{-2} sec^{-1}),
\end{equation}

\begin{equation}
p+p+e^-\to D+\nu_e, \qquad (E_\nu=1.44: \,\phi_\nu=1.5\times
10^8),
\end{equation}
or from the $pp$-II chain,

\begin{equation}
^7Be+e^-\to ^7Li+\nu_e,\qquad (E_\nu=0.86\,MeV:\,
 \phi_\nu=2.7\times 10^9;\quad E_\nu=0.38:\,3.0\times 10^8),
\end{equation}
or from the $pp$-III chain,

\begin{equation}
^8B\to ^8Be*+e^+ +\nu_e,\qquad (E_\nu^{max}=14.06;\,
\phi_\nu=3.0\times 10^6),
\end{equation}
or from the C-N-O cycle,

\begin{equation}
^{13}N \to ^{13}C+e^+ +\nu_e, \qquad (E=1.19: \, 3.0\times 10^8),
\end{equation}

\begin{equation}
^{15}O\to ^{15}N+e^+ +\nu_e, \qquad (E=1.70: \, 2.0\times 10^8).
\end{equation}
Here the neutrino fluxes $\phi_\nu$ are measured at the sea level
on Earth, in units of $cm^{-2}sec^{-1}$. Of course, the
electron-like neutrinos may oscillate into muon-like or tao-like
specifies but fortunately neutral weak interactions do not
differentiate among them; other types of neutrino oscillations, so
far less likely, could be relevant though.

The average distance of the planet Jupiter from the Sun is 5.203
$a.u.$ with the Jupiter year 11.9 our years. The radius of the
Jupiter is 71,398 $km$, much bigger than the Earth's 6,378 $km$.
In terms of the mass, the Jupiter's $1.901\times 10^{27}\,Kg$ is
about 300 times than the Earth's $5.974\times 10^{24}\,Kg$. It is
believed that the composition of the Jupiter is similar to our
Sun, mostly the hydrogen plus a certain fraction of the helium.

Therefore, when solar neutrinos encounter the Jupiter, we
anticipate that the following weak interactions will dominate:
\begin{equation}
\nu + p \to \nu +p, \qquad \nu + ^4He \to \nu + ^4He,
\end{equation}
while the reaction $\nu + e^- \to \nu + e^-$ would serve as a
small correction.

\section{Estimate of the Mean Free Paths}

For the neutral-current weak reaction induced by solar neutrinos
on the protons,
\begin{equation}
\nu(p_\nu)+p(p) \to \nu(p'_\nu)+p(p'),
\end{equation}
the transition amplitude is given by\cite{Hwang}
\begin{equation}
T={G\over \sqrt 2}i {\bar u}_\nu(p'_\nu)\gamma_\lambda(1+\gamma_5)
u_\nu(p_\nu) \cdot <p(p')\mid N_\lambda \mid p(p)>.
\end{equation}
We may proceed to parameterize the neutral-current matrix element
as follows\cite{Hwang}:
\begin{eqnarray}
&<p(p')\mid N_\lambda(0)\mid p(p)> \nonumber\\
=&i\bar u(p') \{\gamma_\lambda f_V^N(q^2)-{\sigma_{\lambda\eta}
q_\eta\over 2m_p}f_M^N(q^2) +\gamma_\lambda \gamma_5 f_A^N (q^2)
+{i2Mq_\lambda \gamma_5\over m_\pi^2}f_P^N(q^2) \}u(p),
\end{eqnarray}
with $q^2\equiv \vec q\,^2-q_0^2$, $q_\lambda=(p'-p)_\lambda$, and
$2M=m_p+m_n$. Here $f_V^N(q^2)$, $f_M^N(q^2)$, $f_A^N(q^2)$, and
$f_P^N(q^2)$, respectively, the (neutral-current) vector, weak
magnetism, axial, and pseudoscalar form factors. The differential
cross section is given by
\begin{eqnarray}
&{d\sigma\over d\Omega_\nu} (\nu + p \to \nu + p) \nonumber\\
=&{G^2(E'_\nu)^2\over 2\pi^2} {E'_\nu \over E_\nu} \{
[(f_V^N(q^2))^2 +(f_M^N(q^2))^2 {q^2\over 4m_p^2} +
(f_A^N(q^2))^2] cos^2 {\theta_\nu\over 2}  \nonumber\\
 & +2[(f_V^N(q^2)+ f_M^N(q^2))^2{q^2\over 4m_p^2}
 +(f_A^N(q^2))^2(1+ {q^2\over 4m_p^2})  \nonumber\\
 & +4{E'_\nu \over m_p}(1+{E_\nu\over m_p}
 sin^2{\theta_\nu\over 2})f_A^N(q^2)
 (f_V^N(q^2)+f_M^N(q^2))]sin^2{\theta_\nu \over 2} \}.
\end{eqnarray}

In the tree approximation in the standard model of particle
physics, we have
\begin{equation}
N_\lambda=(1-2sin^2\theta_W)I_\lambda^3-sin^2\theta_W Y_\lambda
+I_\lambda^{3(5)} - {1\over 2}Y_\lambda^s - {1\over
2}Y_\lambda^{s(5)},
\end{equation}
so that, for example,
\begin{equation}
f_V^N(q^2)=(1-2sin^2\theta_W)\cdot {1\over 2}(e_p(q^2)-e_n(q^2))
-sin^2\theta_W \cdot (e_p(q^2)+ e_n(q^2))-{1\over 2}f_V^S(q^2).
\end{equation}

\begin{equation}
f_M^N(q^2)=(1-2sin^2\theta_W)\cdot {1\over 2}(\mu_p(q^2)
-\mu_n(q^2))-sin^2\theta_W\cdot (\mu_p(q^2)-\mu_n(q^2)) -{1\over
2}f_M^S(q^2).
\end{equation}

\begin{equation}
f_A^N(q^2)={1\over 2}f_A(q^2) -{1\over 2}f_A^S(q^2).
\end{equation}

As a reasonable estimate, we could use $q^2\approx 0$ and neglect
all terms higher order in $q^2/m_p^2$ and $E_\nu/(2m_p)$. The
integration over $d\Omega$ yields

\begin{eqnarray}
\sigma & \cong & {G^2E_\nu^2\over \pi}\cdot \{({\bar f}_V^2+ {\bar
f}_A^2+...)(1+{2E_\nu\over m_p})^{-1}\nonumber\\
&& + (2 {\bar f}_A^2+ ...)(1+{2E_\nu\over m_p})^{-2}\}\nonumber\\
& \approx & 1.686\times 10^{-20}\cdot ({\bar f}_V^2 + 3{\bar
f}_A^2)\cdot ({E_\nu\over 1\,MeV})^2\cdot barn,
\end{eqnarray}
where ${\bar f}_V$ and ${\bar f}_A$ are suitable averages of
$f^N_V(q^2)$ and $f^N_A(q^2)$, respectively.

The neutrinos could come from either the three-body modes (i.e.
the $\beta^+$ decays) or the two-body modes (such as the $\beta^+$
capture reactions). For the three-body modes, we could use the
phase factors to do very good estimates for the neutrino spectra;
we adopt this approximation in this paper.

Our estimate, from Eqs. (1)-(6), for the average flux times the
cross section, $\phi_\nu \sigma$, is given by
\begin{equation}
\phi_\nu\sigma=4.838\times 10^{-36}({\bar f}_V^2+3{\bar f}_A^2)
sec^{-1}.
\end{equation}
The average density of the Jupiter is 1.2469 $gm/cm^3$. The
inverse of the mean free path $n\sigma$ is given by
\begin{equation}
n\sigma=2.102\times 10^{-36} ({\bar f}_V^2+3{\bar f}_A^2) cm^{-1}.
\end{equation}
The neutrino flux suitably weighted by the energy factor, measured
on the surface of the Jupiter, is
\begin{equation}
\phi_\nu=2.869\times 10^8 cm^{-2}sec^{-1}.
\end{equation}
This factor is already used before, calculated from from Eqs.
(1)-(6) adjusted by the distance from the Jupiter and the Sun.

As another estimate, we could compare how much energy the solar
neutrinos deposit in the Jupiter to that in the Earth,
\begin{equation}
({1\over 5.203})^2\times ({71,398 km \over 6,378 km})^3=51.82,
\end{equation}
modulated by small difference in the densities.

Another interesting question to ask: The Sun produces a lot
neutrinos (antineutrinos) per unit of time but we could ask which
stellar object catches the most of them? How many of them get lost
in the empty space? So, the Jupiter does come into play.

\section{Discussions}

Unless the neutrino species would oscillate into the antineutrino
species, the relevant weak interactions are relatively simple in
our problem - only neutral-current weak interactions and, in terms
of the energy range, mostly the elastic channels. So, whether
neutrino-antineutrino oscillation occurs is an important issue.

Neutrino oscillations is now established to be of importance in
the Sun - thus, we could speculate that it is also true in the
Jupiter, the Mini-Sun, a factor of 10 smaller (in diameter). The
scenario for the oscillations is still unknown and yet to be
established, and we feel that the oscillations into antineutrinos
($\Delta L =2$) or into sterile species, if happens, would add a
lot of fun in the game.

Most of these issues can in principle be investigated in
experiments on the Earth - there is no need to go to the Jupiter
to enhance our knowledge. The stopping power of solar neutrinos
may be the only signature, however small it could be (in terms of
the temperature that it increases over a time span). It comes from
the interior of the Jupiter, so it's interesting in this regard.

\bigskip

\section*{Acknowledgments}
The Taiwan CosPA project is funded by the Ministry of Education
(89-N-FA01-1-0 up to 89-N-FA01-1-5) and the National Science
Council (NSC 96-2752-M-002-007-PAE). This research is also
supported in part as another National Science Council project (NSC
96-2112-M-002-023-MY3).

\end{document}